\documentclass[prb,showpacs,twocolumn,superbib,floats,groupedaddress]{revtex4}
\usepackage{graphicx}
\usepackage{amsmath}
\usepackage{bm}

\begin{document}

\title{LDA+U and tight-binding electronic structure of InN nanowires}

\author{A. Molina-S\'{a}nchez, A. Garc\'{i}a-Crist\'{o}bal, A. Cantarero}

\affiliation{Instituto de Ciencia de Materiales, Universidad de
Valencia, E-46071 Valencia, Spain}

\author{A. Terentjevs}

\affiliation{Physics Department, Politecnico of Torino, Torino, Italy}

\author{G. Cicero}

\affiliation{Chemistry and Materials Science Eng. Department, Politecnico of Torino, Torino, Italy}

\date{\today}

\begin{abstract}
In this paper we employ a combined {\it ab initio}  and
tight-binding approach to obtain the electronic and optical
properties of hydrogenated InN nanowires. We first discuss InN band
structure for the wurtzite structure calculated at the LDA+U level
and use this information to extract the parameters needed for an empirical
tight-binging implementation. These parameters are then
employed to calculate the electronic and optical properties of InN
nanowires in a diameter range that would not be affordable by
{\it ab initio} techniques. The reliability of the large nanowires
results is assessed by explicitly comparing the electronic structure
of a small diameter wire studied both at LDA+U and tight-binding
level.
\end{abstract}

\maketitle

\section{Introduction}

Indium nitride (InN) has received considerable attention in recent
years due to its direct bandgap in the infrared
range\cite{nuria1,nuria2} and the high electron
mobilities.\cite{nuria3} The possibility of fabricating
low-dimensional structures such as nanowires
(NWs)\cite{calleja,pal2008} makes desirable the simulation of the
electronic structure and optical properties of these system with
atomistic approaches. \textit{Ab initio}-based calculations are in
principle capable of reproducing the band structure of bulk and
systems of few atoms with a great accuracy. However, the
computational time turns out to be a limiting factor if the number
of atoms increases, making this methods impractical for the study of
dependencies with the size and composition (alloying) of the system.
However, the valuable \textit{ab initio} information can be used for
the development of empirical tight-binding (TB)\cite{carlo} or
pseudopotential\cite{bester} methods. These approaches are expected
to describe with good precision the optical properties of both bulk
and small systems, and at the same time, to allow quantitative
studies of large systems in reasonable computational times.
Moreover, as opposed to the approaches based on the effective mass
approximation (EMA), the empirical atomistic methods are able to
incorporate the true symmetry of the nanostructures.\cite{best1}

\textit{Ab initio} methods have been widely employed for the study
of InN (see e.g. Refs. \onlinecite{walle} and \onlinecite{fuchs}).
In particular it has been shown that in order to get a correct
description of its band structure close to the $\Gamma$ point within
the Density Functional Theory (DFT) it is important to repair the
deficiency of Local Density Approximation (LDA) or Generalized
Gradient Approximation (GGA) functional in describing the Coulomb
interaction between the localized $d$ electrons of Indium. To this
end, various approaches have been built on the DFT basis and, among
others we mention the self-interaction correction
methods\cite{voglABO,vogel} and the LDA with the Hubbard $U$
correction (LDA+U).\cite{anisimov97,cococcioni05} In our work, we
adopt an LDA+U approach, which has been recently discussed and
applied to the case of InN.\cite{alexander} Beyond DFT calculations,
the GW methods provides good estimation of the nitrides band gap,
opening it up the experimental value. However, this method is
computationally more complex than LDA+U, making difficult its
application in large systems.\cite{rinke,svane}

Among the various empirical approaches, we have chosen to work with the
tight-binding method, that has demonstrated its applicability in
III-N nanowires.\cite{camacho} Moreover, this method allows to deal easily
with the problem of the
dangling bonds at the free surface of the nanowires,\cite{delerue} and gives
an intuitive physical picture of the wave functions in terms of the atomic
orbitals. The TB parameters are obtained by fitting the LDA+U bulk band
structure to some selected points of the Brillouin zone, with
special care in a faithful description of the neighborhood of the
top of the valence band, because of its dominant role in the
determination of the optical properties.

Since there is no \textit{a priori} guaranty of the transferability
of the fitted TB parameters for their use in nanostructures
calculations, we have compared the band structure of a small InN
nanowire (diameter 16.2 {\AA}) calculated with the LDA+U and TB
approaches, and obtain a very good agreement. The use of this
empirical tight-binding model in larger InN nanowires has been
illustrated by calculating the dependence of the confinement energy
on the NW size, and the polarization dependence optical spectra for
a nanowire size beyond the range accessible by \textit{ab initio}
calculations.

\section{Indium Nitride Bulk}
\label{bulk}

InN has been studied by employing DFT-LDA\cite{perdew81}
calculations with on-site Hubbard $U$ correction (LDA+U), using
ultra-soft pseudopotentials as realized in Quantum
Espresso,\cite{pwscf} and expanding the electronic wave functions in
plane waves. To describe correctly the structural properties of InN,
the 4$d$ electrons of the Indium are explicitly considered as
valence electrons.\cite{wright95} For all calculations, the plane
wave cutoff is 30~Ry, and a (8$\times$8$\times$8) Monkhorst-Pack
mesh is used.

\begin{figure}[h!]
\includegraphics[width=7.5cm]{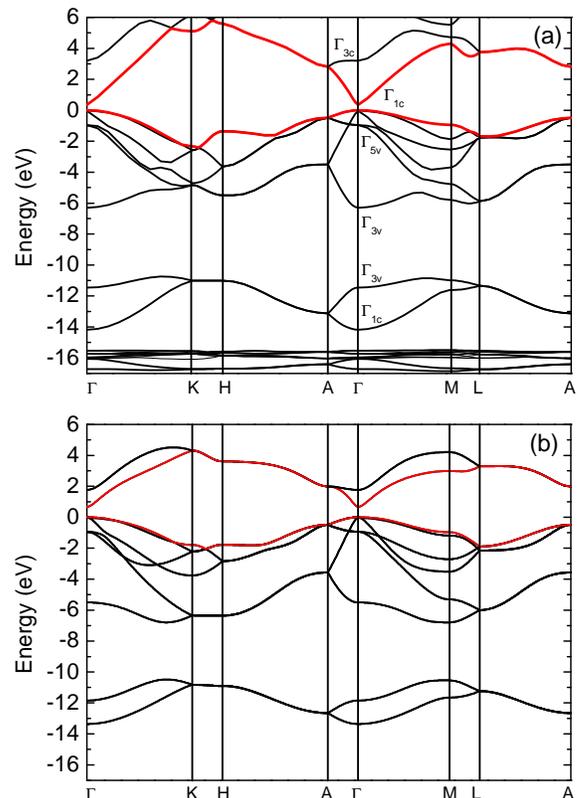}
\caption{(Color online) Band structure of InN bulk obtained with (a)
LDA+U and (b) TB approaches. The symmetry group labels of some
relevant states are indicated in (a).} \label{comp_bulk}
\end{figure}

It is known that LDA and GGA underestimate the binding energy of the
cation semicore $d$ states and overestimate their hybridization with
the anion $p$ valence states. As a result, an artificially large
$p$-$d$ coupling pushes up the valence band maximum reducing the
calculated band gap; in particular, in the case of InN, DFT-LDA
gives null or negative band gap (the experimental band gap is
$\sim$0.67 eV \cite{nuria2,thakur}). In this work we use the
LDA+U\cite{anisimov97,cococcioni05,janotti06} method to correct this
deficiency. To describe correctly the main InN band features, we
have applied the $U$ correction both to indium 4$d$ electrons and to
nitrogen 2$p$ electrons. We note that in the case of InN, similarly
to some oxides compounds,\cite{korotin00} the inclusion of the U
correction on the anion (the N $p$-shell) is important for a better
description of $p$-$d$ interaction and, besides inducing the band
gap opening, it gives the correct symmetry of the states close to
the top of the valence band. The spin-orbit interaction is not taken
into account in these calculations. The selected U parameters are
$U_d$=6.0~eV for In, $U_p$=1.5~eV for N, as discussed in details
elsewhere.\cite{alexander} Within this computational scheme, we
obtained equilibrium lattice parameters for InN bulk in the wurtzite
structure of $a = 3.505$ \AA\ , $c/a = 1.616$, $u = 0.378$. These
values are close to the experimental data ($a = 3.538$ \AA\ , $c/a =
1.612$ , $u = 0.377$ \cite{paczkowicz03}). In Fig.
\ref{comp_bulk}~(a) the band structure for the InN bulk is
presented: the band gap at $\Gamma$ is 0.34 eV, the valence band
width is about 6.3~eV and the 4$d$ indium states lie 16~eV below the
valence band maximum (VBM). The bandgap becomes positive but it is
still underestimated as compared with the experimental value.
Another remarkable improvement with respect to LDA consists of the
correct description of the energy level ordering and symmetry at the
top of the valence band, which are essential to derive reliable TB
parameters.

\begin{table}[h!]
\begin{tabular}{cccccc}
\hline \hline
$\quad E_{s}^c \quad$ & $\quad E_{p}^c \quad$   & $\quad E_{s}^a \quad$ & $\quad E_{p}^a \quad$ & $\quad E_{p_z}^a \quad$ & $\quad \eta \quad$ \\
\hline
-5.5247   & 9.6179    & -6.7910   & 0.0461      &  -0.0076 & 1.8\\
\hline \hline $\quad V_{ss\sigma} \quad$ & $\quad V_{s_cp_a} \quad$
& $\quad V_{s_ap_c} \quad$ & $\quad V_{pp\sigma} \quad$ & $\quad
V_{pp\pi}
\quad$  & $\quad \eta_{s,p_z,\sigma} \quad$ \\
\hline
-1.7500  & 2.5981 & -0.1083 & -1.3000 & 3.0700 & 2.5\\
\hline \hline
\end{tabular}
\caption{TB parameters (in eV) of InN proposed in this work. We
follow the standard TB notation also used in Ref.
\onlinecite{kobayashi}.} \label{table_etb}
\end{table}

Concerning the empirical TB method, we have selected a basis of four
orbital per atom, $s,p_x,p_y,p_z$ ($sp^3$ model), as described in
Ref. \onlinecite{kobayashi}. It is known that a better description of the
conduction bands far from the $\Gamma$ point would require at least the use
of an $sp^3s^*$.\cite{vogl} However, we will focus our study in
the optical properties near $\Gamma$, and to keep the number of fitting
parameters reduced, we avoid the addition
of the $s^*$ excited orbital. As we only include interaction
between nearest neighbors, the crystal-field splitting at the top of
the valence band cannot be reproduced, since this is an effect
caused by the interaction with second and third neighbors. This
limitation is corrected by the introduction of one ad-hoc asymmetry
between $p_x-p_y$ and $p_z$ orbitals.\cite{shulz} Moreover, the
deviation from the ideal wurtzite has been introduced with the
Harrison's rule, applied to the interatomic
parameters:\cite{harrison}

\begin{equation}
V(d)=\left(\frac{d_0}{d}\right)^{\eta}V(d_0),
\end{equation}

where $d$ is the relaxed LDA+U distance, $d_0$ the ideal wurtzite
distance, and $\eta$ an exponent that depends on the orbital. In
most of the literature, the accepted value for the exponent is
around 2,\cite{andres} although some authors make a discretionary
use of such exponents in order to obtain a good description of the
band structure under deformation (see Ref. \onlinecite{boykin}) or a
better agreement over the whole Brillouin zone (see in Ref.
\onlinecite{jancu2002}). In an attempt to limit the number of
additional parameters we restrict the $\eta$ to be different only
for the overlap $s-p_z$. A optimized set of TB parameters fitted
with this procedure against the LDA+U band structure is shown in
Table~\ref{table_etb}, and the corresponding TB band structure is
represented in Fig.~\ref{comp_bulk}~(b). Nevertheless, the bandgap
has been fixed to the experimental value.\cite{nuria1} The obtained
valence band reproduces well the LDA+U results. The discrepancies at
around -6 eV below the top of the valence band are attributed to the
small basis set used in the $sp^3$ TB method.\cite{kobayashi}

\begin{figure}[h!]
\includegraphics[width=8.1 cm]{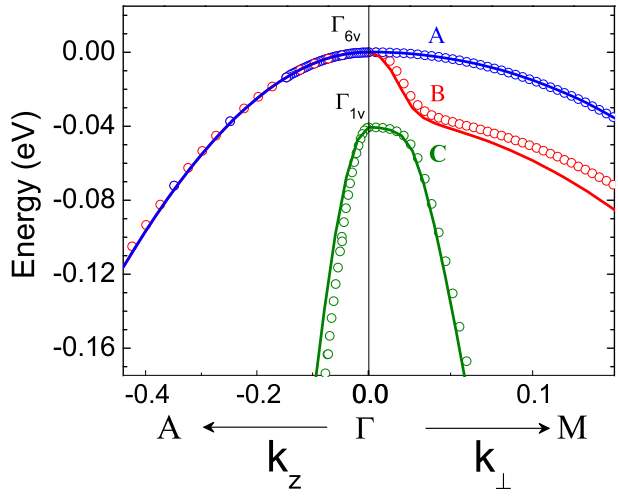}
\caption{(Color online) Top of the InN valence band. The empty
circles correspond to the LDA+U bands, and the lines represent the
TB bands. The component $k_z$ of the wave vector $\bm{k}$ is
normalized to $\frac{\pi}{c}$, such that $k_z=1$ correspond to A.
The wave vector in the $M$ direction is expressed as
$\frac{\pi}{a}(\xi,\frac{1}{\sqrt{3}}\xi,0)$, where $0\leq\xi\leq
1$. The symmetry group of the states at $\Gamma$ are indicated and
the bands are denoted as A, B and C.}\label{zoom_tb}
\end{figure}

In Fig.~\ref{zoom_tb} we report the details of the top of the
valence band at $\Gamma$, comparing LDA+U (open circles) with TB
calculations (lines), along the $\Gamma A$ and $\Gamma M$
directions. We observe a very accurate fitting for the A-C bands,
whereas the B-band shows a slight deviation for $k>0.1$ in the $M$
direction. Along the $\Gamma A$ direction, A and B bands are
degenerate and both calculations match perfectly. The anti-crossing
between B and C bands is also well captured by the TB method. The TB
effective mass of the top of the valence band at $\Gamma$ are
$m_{\bot}^A = 2.80 $, $m_{z}^A = 1.86 $, $m_{\bot}^B = 0.07$,
$m_{z}^B = 1.86 $, $m_{\bot}^C = 0.57 $ and $m_{z}^C = 0.07 $.
Regarding the symmetry of the wave functions at
$\Gamma$,\cite{cardona} the degenerate states belong to the
representation $\Gamma_{6v}$. They have a pure composition of $p_x$
and $p_y$ orbitals, that coincides with both LDA+U and TB results.
The second state in energy belongs to the representation
$\Gamma_{1v}$, being here 100 \% $p_z$ for both calculations. Note
that the bottom of the conduction band state also belongs to this
representation, although the predominant orbital is in this case
$s$-type. The TB conduction effective masses are $m_{\bot}^c=0.07$
and $m_z^c=0.08$, in agreement with the data of Ref.
\onlinecite{wu}. The achieved good agreement at $\Gamma$ is of
special relevance for the eventual use of the TB band structure in
the analysis of optical and transport experiments.

\section{Indium nitride nanowires}
\label{nw}

To asses the behavior of the TB parametrization in nanostructures, a
comparison between the electronic states, calculated with LDA+U and
TB approaches, has been performed, for a thin NW. Afterwards, a
study in larger NWs with the TB method has been carried out, by
exploring the bandgap evolution with the NWs diameter and examining
the optical response for a selected diameter.

\begin{figure*}
\includegraphics[width=15cm]{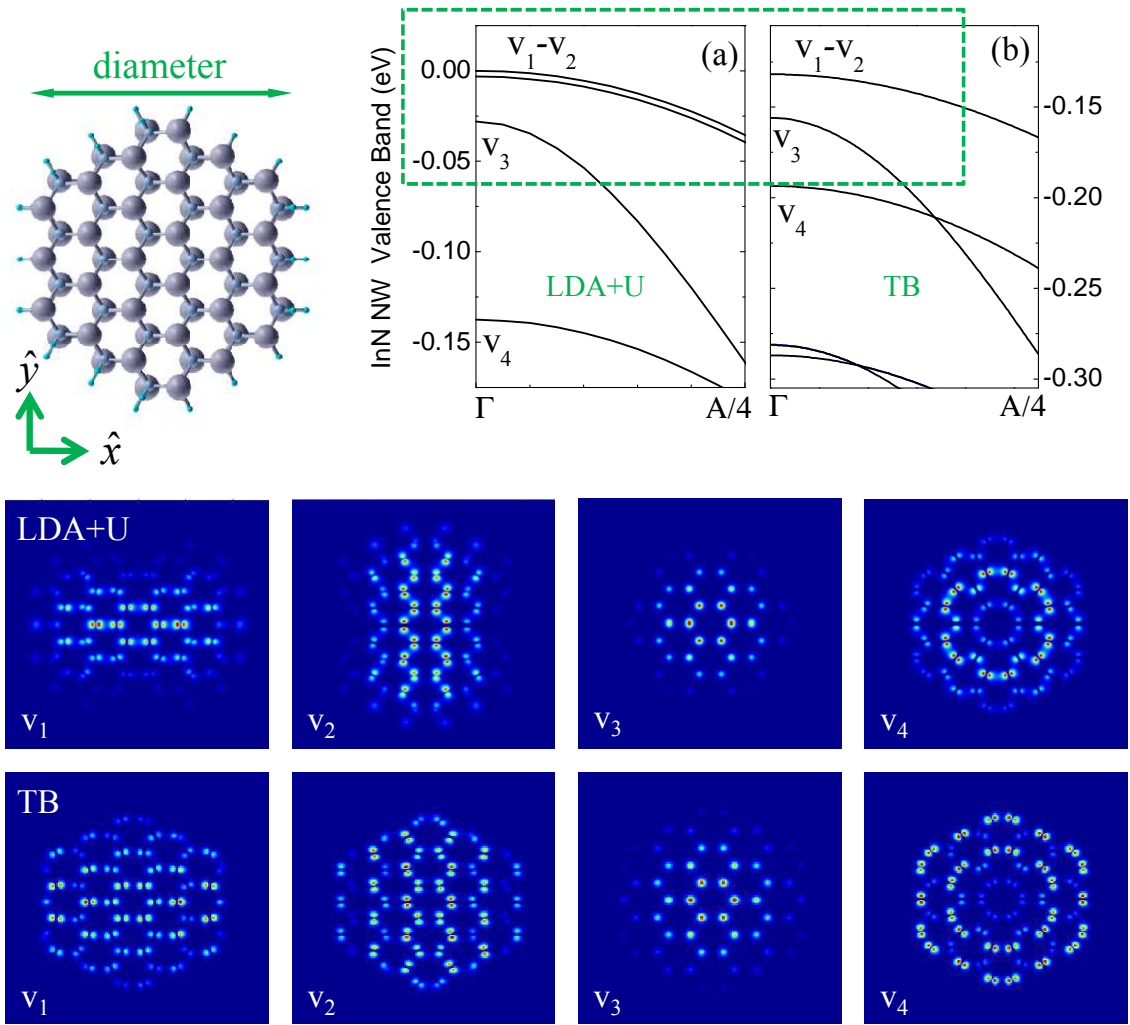}
\caption{(Color online) In the left upper part, nanowire represented
with ball-and-sticks of diameter 16.2 {\AA}. As well in the upper
part, top of the valence band calculated with (a) LDAU (b) and TB
method. In the lower part, we represent the square of the wave
function for the valence band states $v_1$, $v_2$, $v_3$ and $v_4$
calculated with both approaches.} \label{comp_nw}
\end{figure*}

The nanowire employed in the comparison has a diameter of 16.2 {\AA}
(see sketch in the left upper part of Fig. \ref{comp_nw}), and the
dangling bonds at the free surfaces are passivated with hydrogen
atoms in order to avoid the presence of surface states within the
gap.\cite{huang} In the \textit{ab initio} calculation, the nanowire
structure has been fully optimized until forces on atoms are less
than 0.001 Ry/bohr per atom. We use a Monkhorst-Pack mesh of 6
points for the one dimensional nanowire Brillouin zone. The indium
and nitrogen atoms placed at the surface modify slightly its
tetragonal bond due to the presence of the passivant hydrogen atoms,
changing slightly their interatomic distances. This surface
reconstruction is not taken into account in the TB calculation,
which assumes a perfect wurtzite everywhere.\cite{persson} The
topmost valence band states, labeled in increasing energy as
$v_1,v_2,...$, are shown in Fig. \ref{comp_nw} ((a) LDA+U and (b) TB
method). The states $v_1$ to $v_4$ are within a range of 150 meV in
both calculation. The TB result yields in addition the value of -130
meV for the confinement energy of the states $v_1$ and $v_2$ with
respect to the top of bulk valence band. In the case of the LDA+U
calculation, the degeneracy between $v_1$ and $v_2$ is broken due to
the exact consideration of the atomic distances when we relax the
structure, an effect that the TB method ignores (such splitting has
the small value of 3 meV). In any case, the portion of the band
structure framed by a dashed green line, that contains the $v_1$,
$v_2$, and $v_3$ sub-bands, exhibit a remarkable similarity in both
calculations. In particular, the curvature of the bands are
identical and only a slight difference between the $v_1$ and $v_3$
states (26 meV and 18 meV for LDA+U and TB calculation,
respectively) is observed. Concerning the $v_4$ state, one can
perceive that is closer in energy to $v_3$ in the TB calculations
than in the LDA+U approximation. Despite that energy is
underestimated, $v_4$ has the same curvature in both approaches.
Another difference between both methods is the existence of more
states in the range of -150 meV from the state $v_1$, in the case of
TB valence band. In order to exclude the relaxation as a source of
error in our comparison, calculations with LDA+U in a nanowire,
assuming perfect wurtzite everywhere were performed, without finding
any substantial difference.

In the lower part of Fig.~\ref{comp_nw} we show the square of the
$\Gamma$ wave function, $<\Psi|\Psi>$, for the valence band states,
$v_1$ to $v_4$. The TB wave function can be expressed as:

\begin{equation}
\Psi_{\Gamma}(\bm{r}) =
\sum_{\alpha,j}A_{\alpha,j}\phi_j(\bm{r}-\bm{r_{\alpha}}),
\end{equation}

here the index $\alpha$ runs over atoms and $j$ over orbitals. For
the sake of simplicity, the orbitals are represented here with the
hydrogen wave functions that share the same symmetry.\cite{galindo}
Figure~\ref{comp_nw} shows that the density is localized on the
indium and nitrogen atoms, without spreading on the hydrogen atoms.
The first two degenerate valence band states ($v_1$ and $v_2$)
exhibit the electron density elongated along two perpendicular
directions ($x$ and $y$). Moreover, by looking closely to the
density of each atom, it is evident that it comes from the $p_x$ and
$p_y$ orbitals, for the $x$-elongated ($v_1$) and $y$-elongated
($v_2$) states, respectively. In the next valence band state, $v_3$,
the wave function is notably confined at the center of the NW, being
the $p_z$-orbital component predominant. In the case of the $v_4$
state, we find that the wave function has a node in the nanowire
center, and a mixed composition of $p_x-p_y$ orbitals. One can
distinguish that TB charge densities are more delocalized towards
the NW surface if compared to the LDA+U picture. Even so, TB method
reproduces exactly the qualitative features of the charge density in
terms of symmetry and orbital composition. The observed differences
are acceptable because of the restricted TB basis and the small
sizes of the NW. For larger NWs, these small differences between the
TB method and the LDA+U approximation are expected to be attenuated.
We thus conclude that the TB parameters obtained and tested here are
suitable to be used in the calculation of optical properties of
InN-based nanostructures.

\begin{figure}[h!]
\includegraphics[width=\linewidth]{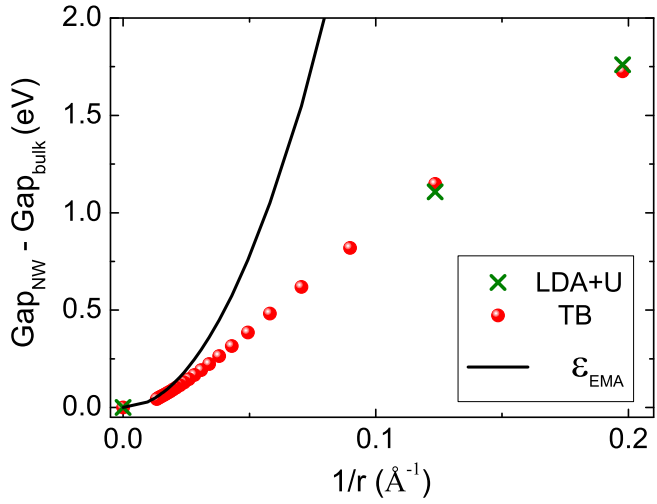}
\caption{\label{gap_size}(Color online) Dependence of the
confinement energy on the nanowire size. The limit of $1/r
\rightarrow 0$ is the bulk bandgap.}
\end{figure}

Once demonstrated the reliability of the TB approach and the quality
of the parameters, we have performed TB calculations for larger NWs.
In the first place, we show in Fig.~\ref{gap_size} the confinement
energy, defined as the difference between the nanowire and bulk
bandgap, versus $1/r$, being $r$ the NW radius.  The full circles
correspond to the TB results and the bandgap energies calculated
with LDA+U for two NWs are drawn with full rectangles. The
confinement energy calculated with the effective mass approximation,
assuming parabolic bands, is:

\begin{equation}
\varepsilon_{\rm EMA} =
\left(\frac{\hbar^2}{2m_{\bot}^c}+\frac{\hbar^2}{2m_{\bot}^A}\right)\left(\frac{k^0_1}{r}\right)^2,
\end{equation}

$k^0_1=2.4048$, being the first zero of the Bessel function
$J_0(x)$, and the effective masses are reported in Sec.~\ref{bulk}.
For large radii, when $1/r < 0.03$~{\AA}$^{-1}$, the TB method and
the EMA follow the same trend, proportional to $1/r^2$. For
decreasing radii ($1/r > 0.03$~{\AA}$^{-1}$) EMA overestimate the
confinement energy as compared with the TB results, that changes in
this range the $\sim 1/r^2$ behavior to $\sim 1/r$. Moreover, the TB
results connect perfectly with the \textit{ab initio} computed
values, represented by full squares, at radii 8.1~{\AA} and
5.1~{\AA}. This smooth interpolation confirms the suitability of the
TB method to link the NWs size ranges of 10~{\AA}, where \textit{ab
initio} are practical and 100~{\AA}, where the (EMA) start to be
applicable. In this intermediate size range the TB approach has the
advantages of keeping the atomistic nature of the system and be
efficient in terms of computational effort.

In addition the TB method offers the possibility of calculating the
optical absorption spectra without introducing new parameters in the
model. The absorption coefficient for light with polarization vector
$\bm{e}$ can be written as:\cite{grosso}

\begin{equation}
\alpha^{\bm{e}}(\hbar\omega)\propto\int_{BZ}f^{\bm{e}}_{c,v}(k)\delta(E_{c,k}-E_{v,k}-\hbar\omega),
\end{equation}

where we integrate over the one-dimensional Brillouin zone, and the
oscillator strength is calculated as:

\begin{equation}
f^{\bm{e}}_{c,v}(k)\propto\frac{|<\Psi_c|\bm{e}\cdot\bm{p}|\Psi_v>|^2}{E_{c,k}-E_{v,k}}.
\end{equation}

\begin{figure}
\includegraphics[width=0.6\linewidth]{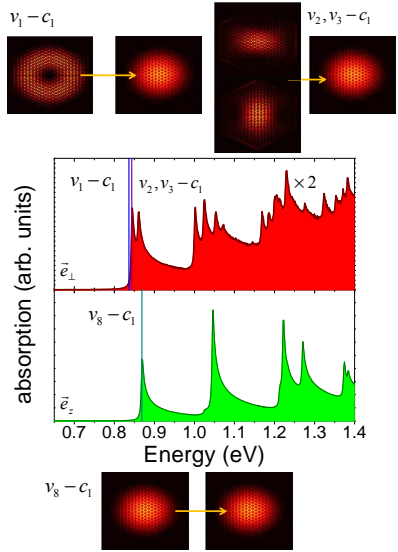}
\caption{(Color online) Optical absorption spectra for in-plane
($\bm{e_{\bot}}$ multiplied by two) and on-axis ($\bm{e_{z}}$) light
polarization (see main text). The wave functions that participate in
the relevant optical transitions are also represented (both spectra
are displayed in the same scale).} \label{opt}
\end{figure}

The momentum matrix element, $<\Psi_c|\bm{e}\cdot\bm{p}|\Psi_v>$, is
calculated as in Ref.~\onlinecite{rammohan}, for two light
polarizations: in-plane (perpendicular to NW axis),
$\bm{e}_{\bot}=1/\sqrt{2}(\hat{x}+i\hat{y})$, and in-axis (parallel
to NW axis), $\bm{e}_z=\hat{z}$. The delta function is replaced by a
Lorentz function of width 7 meV. In Fig.~\ref{opt} we represent the
absorption spectrum of a NW of diameter 70.8~\AA\ for the defined
light polarizations. The valence and conduction wave functions that
participate in the transitions at the absorption edge are also
shown. In both spectra it is recognized the one-dimensional density
of states (modulated by the oscillator strength), the $\bm{e_{z}}$
spectra exhibiting a larger separation between the absorption peaks.
By analyzing more in more detail the $\bm{e_{\bot}}$ spectra, one
can appreciate that absorption edge does not take place at the
energy of the fundamental bandgap (corresponding to the transition
$v_1-c_1$). This is because the symmetry valence state $v_1$ (see
Fig. \ref{opt}), whose charge density has a node in the at the NW
center, making negligible the spatial overlap between the states
$v_1$ and $c_1$. The first optically active transition, blue-shifted
10 meV with respect to the fundamental gap, involves the degenerate
states $v_2$ and $v_3$, shown in Fig. \ref{opt}. On the other hand,
the absorption edge of the $\bm{e_{z}}$ spectra is shifted 24 meV
with respect the $\bm{e_{\bot}}$ spectra, since the first state with
significant $p_z$ orbital component is $v_8$.

\section{Conclusions}
\label{conclusions}

In this work, we have obtained an InN band structure with a
fundamental bandgap of 0.34 eV, by means of LDA+U calculations. The
Hubbard $U$ correction to the $d$ orbitals of indium and $p$
orbitals of nitrogen has palliated the zero bandgap problem of InN,
present in LDA or GGA calculations. The LDA+U band structure has
been fitted with a $sp^3$ tight-binding model obtaining a very
reasonable overall agreement despite the small size of the TB basis.
It is specially noticeable the satisfactory coincidence between the
energy and symmetry of the wave functions at $\Gamma$ point.

This fitted set of TB parameters is in principle usable for
calculations of the electronic structure of quantum wells, wires
or/and dots. In order to test the suitability of this empirical
approach, a band structure calculation is performed of a InN NW of
16.2 {\AA} diameter and compare with the corresponding LDA+U
calculation, which includes a previous relaxation of the atomic
positions. This comparison shows that, without any additional
fitting, the TB band structure and wave functions matches adequately
with their \textit{ab initio} counterparts. Possibly, the remaining
differences between the two models could be reduced by employing a
TB model with an extended orbital basis set, although this would
increase the number of parameters and computational time. The study
the evolution of the NW bandgap with the radius confirms the
adequacy of TB method to connect efficiently very small sizes
nanoobjects (a few {\AA}) accessible with \textit{ab initio}
approaches, with large sizes nanostructures (hundreds of {\AA}),
where continuous methods are commonly employed. Finally, the
potential of this empirical atomistic approach is illustrated by the
analysis of the absorption of a large nanowire.

\section{Acknowledgments}
This work has been supported by the the NANOLICHT project
(NanoSci-ERA) and the Ministry of Science and Innovation
(MAT2009-10350). Computer time was provided by CINECA through the
CNR-INFM "Iniziativa Calcolo Parallelo" and by Tirant Supercomputer
of the Red Espa\~{n}ola de Supercomputaci\'{o}n (RES), hosted in the
University of Valencia.

\end{document}